\documentclass[aps,prd,letterpaper,twocolumn,preprintnumbers,floatfix,superscriptaddress]{revtex4}

\usepackage{graphicx}
\usepackage{dcolumn}
\usepackage{bm}
\usepackage{epsfig,cancel,color}
\usepackage{amsmath}
\usepackage{amssymb}

\newcommand{\be}{\begin{equation}}
\newcommand{\ee}{\end{equation}}
\newcommand{\beqn}{\begin{eqnarray}}
\newcommand{\eeqn}{\end{eqnarray}}

\newcommand{\GeV}      {~\mathrm{GeV}}
\newcommand{\TeV}      {~\mathrm{TeV}}

\def\beqn{\begin{eqnarray}}
\def\eeqn{\end{eqnarray}}
\def\beqs{\begin{subequations}}
\def\eeqs{\end{subequations}}
\def\beq{\begin{equation}}
\def\eeq{\end{equation}}
\def\ba{\begin{array}}
\def\ea{\end{array}}

\newcommand{\non}{\nonumber \\}

\newcommand{\mathsym}[1]{{}}

\def\mL{\mathcal{L}}
\def\mM{\mathcal{M}}

\def\mO{\mathcal{O}}

\def\mR{\mathcal{R}}

\def\mU{\mathcal{U}}
\def\mV{\mathcal{V}}

\input epsf

\def\hf{\frac{1}{2}}

\usepackage{ulem,fancyvrb}

\begin{document}
\title{The cancellation mechanism in the predictions of electric dipole moments}

\author{Ligong Bian}
\email{lgbycl@cqu.edu.cn}
\affiliation{ Department of Physics,
Chongqing University, Chongqing 401331, China}
\author{Ning Chen}
\email{ustc0204.chenning@gmail.com}
\address{Department of Physics, University of Science and Technology Beijing, 
Beijing 100083, China}

\begin{abstract}

The interpretation of the baryon asymmetry of the Universe necessitates the CP violation beyond the Standard Model (SM). 
We present a general cancellation mechanism in the theoretical predictions of the electron electric dipole moments (EDM), quark chromo-EDMs, and Weinberg operators.
A relative large CP violation in the Higgs sector is allowed by the current electron EDM constraint released by the ACME collaboration in 2013, and the recent $^{199}$Hg EDM experiment. 
The cancellation mechanism can be induced by the mass splitting of heavy Higgs bosons around $\sim\mO(0.1-1)$ GeV, and the extent of the mass degeneracy determines the magnitude of the CP-violating phase. 
We explicate this point by investigating the CP-violating two-Higgs-doublet model and the minimal supersymmetric Standard Model. 
The cancellation mechanism is general when there are CP violation and mixing in the Higgs sector of new physics models.
The CP-violating phases in this scenario can be excluded or detected by the projected $^{225}$Ra EDM experiments with precision reaching $\sim10^{-28}\,e\cdot{\rm cm}$, as well as the future colliders.

\end{abstract}


\maketitle


\noindent{\bfseries Introduction}
\label{section:intro}
The explanation of one of the most fascinating questions in physics, the baryon asymmetry of the Universe (BAU)~\cite{Ade:2013zuv,Beringer:1900zz}, requires the CP violation (CPV) beyond the Standard Model (SM), as observed by Sakharov~\cite{Sakharov:1967dj}.  
By now, the LHC does not obtain sufficient sensitivities of measuring the CP properties of the SM-like Higgs boson directly.
On the other hand, the fast progress in the indirect detection of CPV sets far more stringent limits on CPV beyond SM in comparison with that of LHC, which include the recent electric dipole moment (EDM) measurements of the electrons~\cite{Baron:2013eja}, neutrons~\cite{Baker:2006ts}, and diamagnetic atoms of $^{199}$Hg~\cite{Graner:2016ses} and $^{225}$Ra~\cite{Parker:2015yka}. 
The observation of a non-zero EDM in the near future may be a signature of new physics (NP) needed to account for the BAU puzzle.

The testability makes the electroweak baryogenesis (EWBG) a most popular and attractive mechanism to explain the BAU puzzle~\cite{Morrissey:2012db,Arkani-Hamed:2015vfh}.
One key ingredient of EWBG mechanism is the strong first order phase transition (SFOPT) that helps to understand the electroweak symmetry breaking, and may be tested by the future high energy proton-proton collider~\cite{Arkani-Hamed:2015vfh}. 
The other essential ingredient is the CPV beyond the SM, which would be probed by EDM measurements. 
As studied previously in Refs.~\cite{Shu:2013uua,Bian:2014zka}, the cancellation of Barr-Zee diagrams validates some parameter spaces from the bounds of eEDM measurements released by the ACME collaboration~\cite{Baron:2013eja}, and a large CPV magnitude in the SM-like Higgs sector to explain the BAU via the EWBG mechanism is allowed. 
Meanwhile, the CPV magnitude in the SM-like Higgs sector can be highly constrained with the increasing precision of the EDM measurements as explored in Ref.~\cite{Inoue:2014nva}. 
Therefore, it is necessary to study the CPV beyond the SM, as well as more general cancellation mechanism to avoid the current bounds from EDM measurements. 
As we have checked, the recent $^{199}$Hg EDM measurement~\cite{Graner:2016ses} excludes the benchmark scenario of MSSM being explored in Ref.~\cite{Bian:2014zka} that was valid to implement the EWBG.

This paper presents a more general cancellation mechanism. 
We explore the issue in the framework of the CPV two-Higgs-doublet model (2HDM) and the minimal supersymmetric Standard Model (MSSM). 
We find that the mass degeneracies of the heavy Higgs bosons play a key role in such cancellation mechanism. 
The EDMs in the case of a relatively larger mass splitting of heavy Higgs bosons are almost induced by the CPV mixing between the SM-like Higgs boson and the heavy Higgs bosons, as will be illustrated in the type-II CPV 2HDM. 
In this situation, the cancellation is mainly driven by the SM-like Higgs mediated Barr-Zee diagrams.
For the nearly degenerate heavy Higgs bosons scenario, the cancellations of EDMs are mostly driven by the cancellation of Barr-Zee diagrams of the fermion EDM (fEDM) and the CEDM, as well as the Weinberg operators.

We point out that, the current eEDM and the recent $^{199}$Hg EDM measurements may not be sensitive to probe the CPV in the scenario where such cancellation happens. 
the future precise measurements of the $^{225}$Ra EDM up to $\sim 10^{-28}\,e\cdot{\rm cm}$ turn out to be more powerful to constrain or detect the CPV effects for this scenario. 
The cancellation in the theoretical predictions of EDMs reopen some parameter space in the CPV models, therefore validate the EWBG to explain BAU.


\noindent{\bfseries General analysis}
\label{section:general}
The CPV can stem from the mixings between the $CP$-even and $CP$-odd states in the Higgs sectors,
\beqn\label{eq:Hmatrix}
\mM^2=\left( \ba{cc} M_H^2 &  M_{HA}^2 \\   M_{HA}^2  &  M_A^2  \ea  \right)\,,
\eeqn
with $M_H^2$, $M_A^2$, and $ M_{HA}^2$ being mass squared matrices of the $CP$-even Higgs bosons, $CP$-odd Higgs boson, and the mixing terms between the two. 
When the $M_{HA}^2$ arise from the tree level, the SM-like Higgs boson can either take the CPV couplings, or decouple from the CPV sources. 
A typical example is the CPV 2HDM.
For the mixings stemming from the loop level, one finds that the CPV almost decouples from the SM-like Higgs sector, such as the MSSM case.

\begin{figure}[htb]
\includegraphics[width=.95\columnwidth]{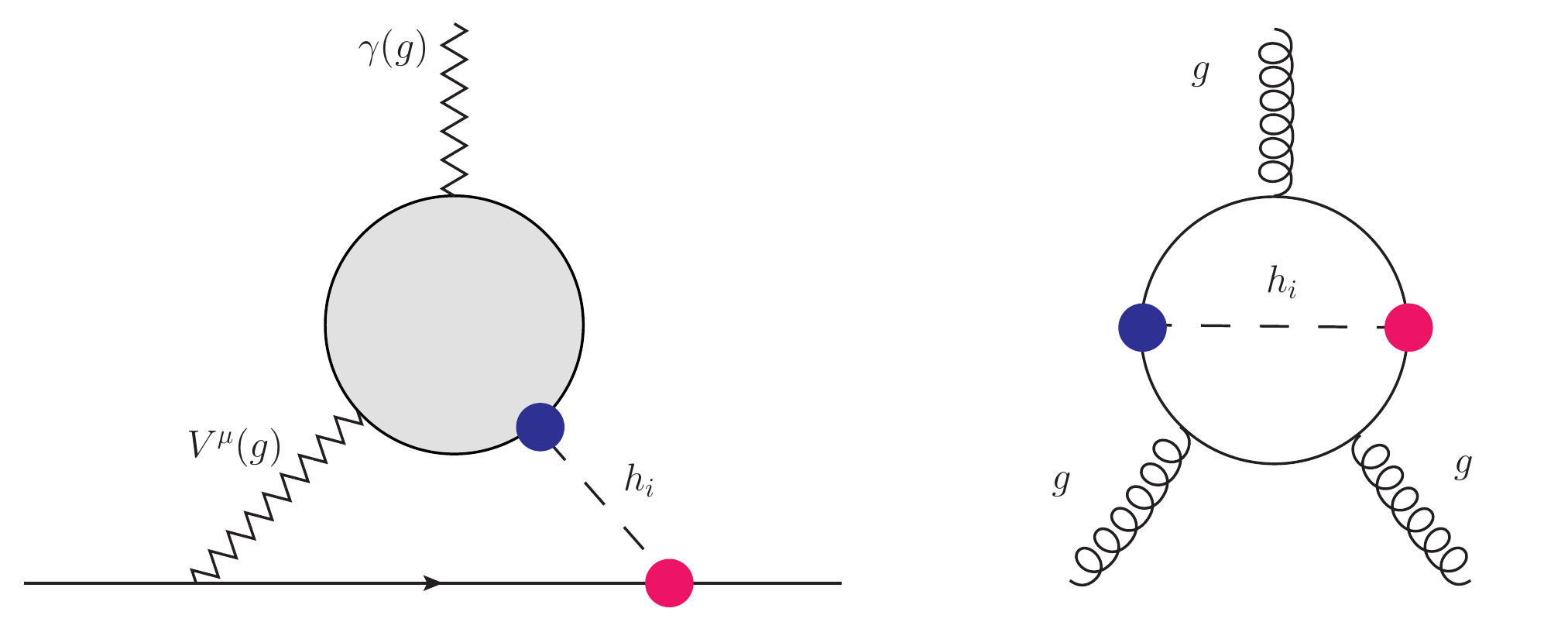}
\caption{\label{fig:BZ_dia}
Left: Two-loop Barr-Zee diagrams for e(q)EDM and CEDM.
Right: Feynman diagram for the Weinberg operators. 
$h_i$ represent all neutral CPV Higgs bosons in Eq.~\eqref{eq:Hmatrix}.
The red (blue) blobs represent the $CP$-even (odd) couplings of Higgs bosons.
}
\end{figure}

The mixings between the $CP$-even and $ CP$-odd Higgs bosons lead to the effective Lagrangian of
\beqn
\mL_{\rm eff}&=& \frac{m_f}{v} h_i \bar f ( c_f^i + i \tilde c_f^i \gamma_5 ) f\non
&+& \frac{\alpha}{\pi v} h_i \Big( c_{\gamma}^i F^{\mu\nu} V_{\mu\nu} + \tilde c_{\gamma}^i F^{\mu\nu} \tilde V_{\mu\nu} \Big)\non
&+& \frac{\alpha_s }{12\pi v} h_i \Big( c_g^i G_{\mu\nu}^a G^{\mu\nu\,,a} + \tilde c_g^i G_{\mu\nu}^a \tilde G^{\mu\nu\,,a}  \Big)\,,
\eeqn
with $(F_{\mu\nu}\,, V_{\mu\nu}\,,G_{\mu\nu}^a)$ being the field strengths of photons, photons and $Z$ bosons, and gluons. 
$c_f^i$ and $\tilde c_f^i$ represent the dimensionless $CP$-even and $CP$-odd Yukawa couplings, and $(c_{\gamma}\,, \tilde c_{\gamma}\,, c_g\,,\tilde c_g)$ are the corresponding Wilson coefficients of the dimension-five operators.
The contributions to the fEDM and CEDM mostly come from Barr-Zee diagrams (the left panel of Fig.~\ref{fig:BZ_dia}). 
After integrating out the internal fermionic and bosonic degrees of freedom, they read
\beqs
\beqn
&&d_{f}^E\sim - c_{f}^i \tilde c_{\gamma}^i \log \frac{ \tilde \Lambda_{\rm UV}^2}{M_i^2}+ \tilde c_{f}^i c_{\gamma}^i \log \frac{ \Lambda_{\rm UV}^2}{M_i^2} \,,\label{eq:eqEDM}\\
&&d_q^C\sim - c_q^i \tilde c_g^i \log \frac{\tilde \Lambda_{\mU\mV}^2 }{M_i^2} + \tilde c_q^i c_g^i \log \frac{\Lambda_{\mU\mV}^2 }{M_i^2}  \, .\label{eq:CEDM}
\eeqn
\eeqs
Here, $\Lambda_{\rm UV}\, (\tilde \Lambda_{\rm UV})$ and $\Lambda_{\rm \mU\mV}\, (\tilde \Lambda_{\rm \mU\mV})$ are relevant scales for the $h_i F^{\mu\nu} V_{\mu\nu}\, (h_i F^{\mu\nu} \tilde V_{\mu\nu})$ and $h_i G^{\mu\nu\,,a} G_{\mu\nu}^a\, (h_i G^{\mu\nu\,,a} \tilde G_{\mu\nu}^a)$ operators. 
Currently, the most stringent constraint on eEDM comes from the ACME, which severely bounds the magnitude of CPV, unless cancellation happens (see Ref.~\cite{Bian:2014zka} for details).
The mixings between $CP$-even and $CP$-odd Higgs bosons also induce dimension-six Weinberg operator (the right panel of Fig.~\ref{fig:BZ_dia}), with the corresponding Wilson coefficient being
\beqn\label{eq:CG}
C_G&\sim& c^i_q \tilde{c}^i_q\, \log \frac{ \Lambda_{\mU\mV}^2 }{M_i^2}\,.
\eeqn
The predictions of the nEDM and diamagnetic aEDMs are determined by all three contributions of qEDM, CEDM and Weinberg operators~\cite{Engel:2013lsa}.
Hence, the nEDM and aEDMs are likely to be suppressed by the cancellations inside two terms in Eqs.~\eqref{eq:eqEDM} and~\eqref{eq:CEDM}, or among different Higgs contributions in Eqs.~\eqref{eq:eqEDM}, \eqref{eq:CEDM}, and \eqref{eq:CG}. 
This in turn, may allow a relatively large CPV in the Higgs sectors needed by the EWBG.

\noindent{\bfseries Type-II 2HDM} 
\label{section:2HDM}
The 2HDM was motivated to offer extra CPV sources from the scalar sector~\cite{Lee:1974jb}, which reside in two following terms with complex coefficients in the Higgs potential
\beqn
V_{\rm 2HDM}&\supset& - m_{12}^2 (\Phi_1^\dag \Phi_2) + \hf \lambda_5 (\Phi_1^\dag \Phi_2)^2 + H.c.\,.
\eeqn
The ratio between two Higgs VEVs is parametrized as $t_\beta\equiv v_2/v_1$, and we denote the real component of the mass mixing term as $m_{\rm soft}^2\equiv {\rm Re}(m_{12}^2)$. 
The imaginary components of $m_{12}^2$ and $\lambda_5$ are sources of CPV, which lead to the mixings among three neutral states as $(h_1\,,h_2\,,h_3)^T= \mR\, (H_1^0\,, H_2^0\,,A^0)^T$. 
The CPV mixing angle between the light Higgs and heavy $CP$-odd Higgs is $\alpha_b$, and the mixing angle between the two heavy $CP$-even and $CP$-odd Higgs bosons is $\alpha_c$.
We always assume $h_1$ being the SM-like Higgs boson with mass of $126\,\GeV$, and $(h_2\,,h_3)$ being two heavy neutral Higgs bosons~\footnote{The masses for the neutral Higgs bosons are denoted by $(M_1\,,M_2\,,M_3)$ in the context.}.
The mixing matrix of $\mR$ diagonalizes the neutral mass matrix as $\mR\,\mM_0^2\, \mR^T={\rm diag}(M_1^2 \,,M_2^2\,, M_3^2)$, and its explicit expression can be found in Ref.~\cite{Khater:2003wq}.   
We focus on the alignment limit of $\beta-\alpha=\pi/2$.
The corresponding Higgs gauge and Yukawa couplings can be found in Refs.~\cite{Bian:2014zka,Inoue:2014nva}.

The cancellation can occur in the evaluation of eEDM when the CPV decouples from the heavy Higgs sector, and the details can be found in Ref.~\cite{Bian:2014zka}~\footnote{ The Ref.\cite{Hayashi:1994ha} found that the destructive contributions of two Higgs bosons would reduce the eEDM and nEDM a lot in the case of the maximal CPV. 
}.
Below, we will explore the complementary case with the CPV almost existing in the heavy Higgs sector, i.e., $|\alpha_b| \ll |\alpha_c|$.
This highly depends on the mass splitting between two heavy Higgs bosons of $\Delta M\equiv M_{3}-M_{2}$ and $t_\beta$, with the following relation~\cite{Khater:2003wq} maintained~\footnote{
We mention two points here. 
First, one can get back to the $t_\alpha=-1$ solution when the mass splitting of $\Delta M=0$, with the related phenomena explored in Ref.~\cite{Bian:2016awe}. 
$\alpha_c$ is unphysical in this situation, and the cancellation mechanism investigated in Ref.~\cite{Jung:2013hka} does not apply. 
Second, $\alpha_b$ cannot be zero, which will otherwise restore to the CP conserving case, as indicated by Eq.~\eqref{eq:Hmass_relation}.}\footnote{The CP-violating mixing can also be enhanced by the possibility of
degeneracies in the SM-like Higgs sectors, as explored in Ref.\cite{McKeen:2012av} with the assistance of singlet and vector-like fermions.
}, 
\beqn\label{eq:Hmass_relation}
&&( M_1^2 - M_2^2 s_{\alpha_c}^2  - M_3^2 c_{\alpha_c}^2 ) s_{\alpha_b} (1 + t_\alpha)\nonumber\\
&&= (M_2^2 - M_3^2 ) (t_\alpha t_\beta - 1) s_{\alpha_c} c_{\alpha_c}\,.
\eeqn

\begin{figure}[htb]
\includegraphics[width=3.9cm,height=3.cm]{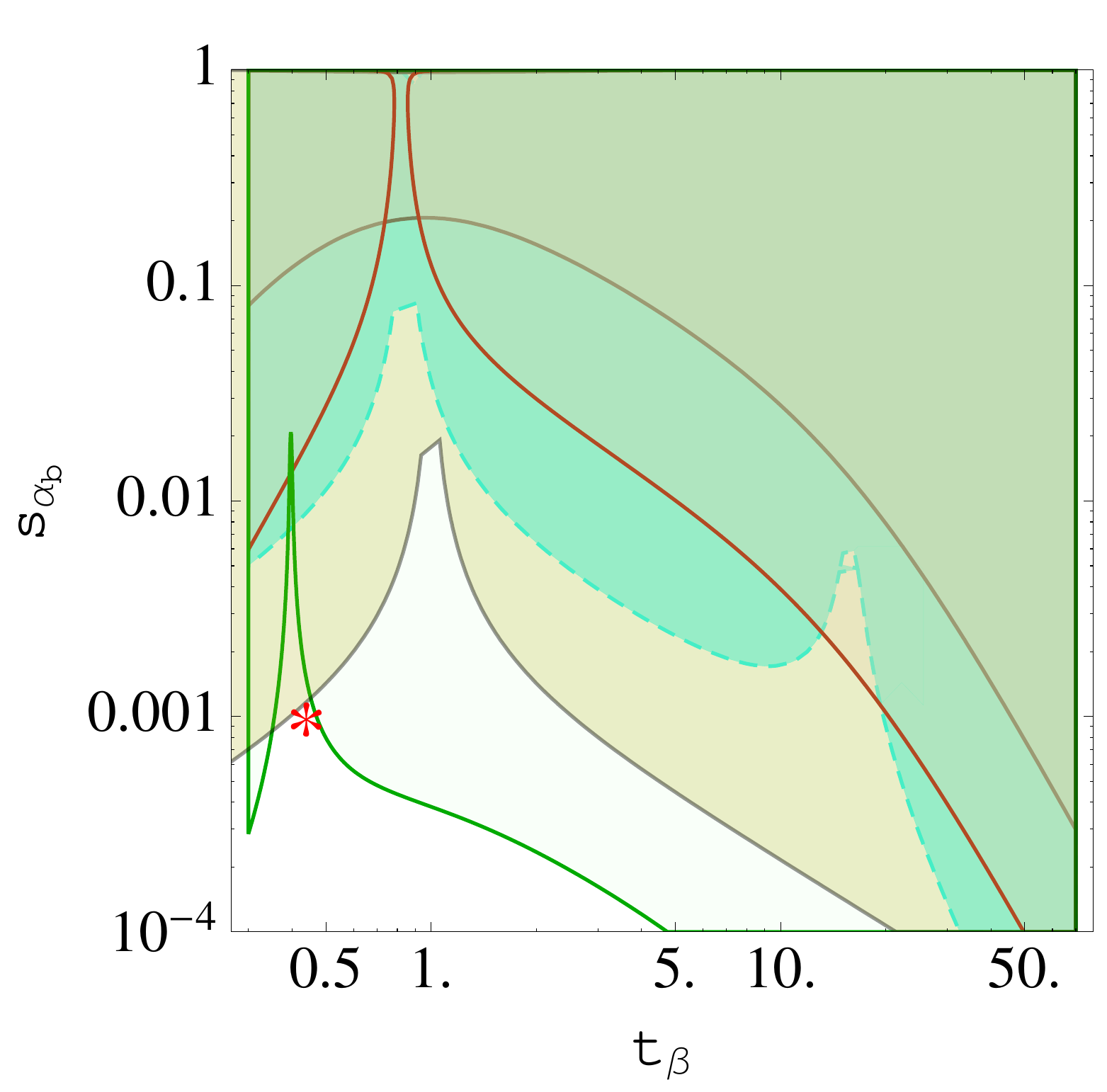}
\includegraphics[width=3.9cm,height=3.cm]{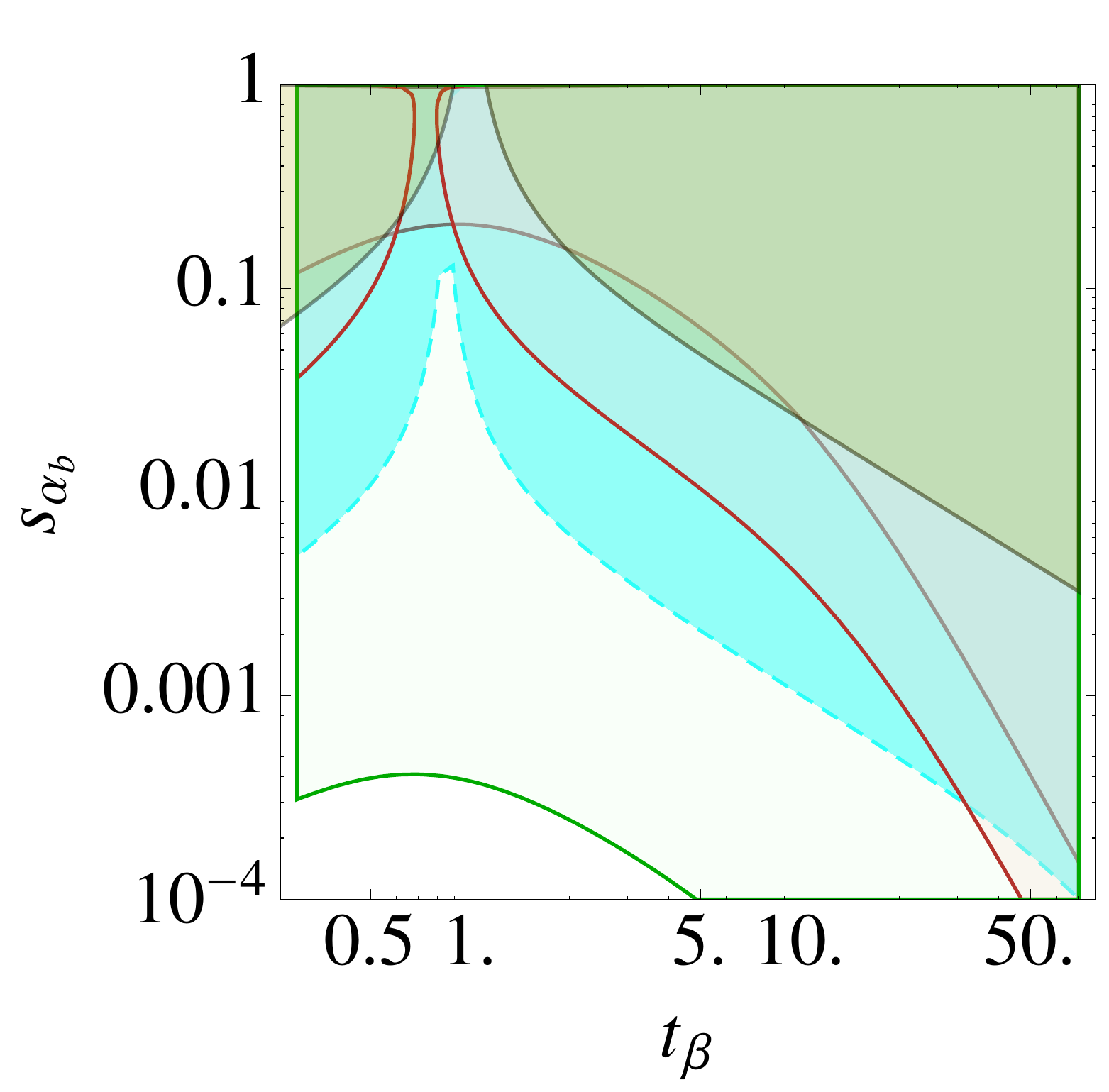}
\caption{\label{fig:BZ}
The EDM constraints on the $(t_\beta\,, s_{\alpha_b})$ plane, with $M_3\,(m_{\rm soft})=500\,(300)\,\GeV$, $\Delta M$ being $0.5\,\GeV$ (left), and $50\,\GeV$ (right).
The EDM experimental exclusions are: $|d_e/e|<8.7\times10^{-29} {\rm cm}$ (Cyan)~\cite{Baron:2013eja}, $|d_n/e|<2.9\times 10^{-26} {\rm cm }$ (Gray)~\cite{Baker:2006ts}, $|d_{\rm Hg}/e|<7.4 \times 10^{-30} {\rm cm}$ (Red)~\cite{Graner:2016ses}, and the projected $|d_{\rm Ra}/e|<1.0\times10^{-28} {\rm cm}$ (Light Green)~\cite{Bishof:2016uqx}. 
The yellow region is theoretically inaccessible since Eq.~\eqref{eq:Hmass_relation} does not have a real solution.}
\end{figure}

To demonstrate the relation between the specific type of cancellation mechanism and the mass splitting between two heavy Higgs bosons, we display EDMs with $\Delta M=0.5$ GeV and $\Delta M=50$ GeV.
The dominant contributions to the e(q)EDMs and CEDM come from the $h_i$-mediated Barr-Zee diagrams~\cite{Barr:1990vd}.
In the case of $\Delta M=0.5$ GeV (the left panel of Fig.~\ref{fig:BZ}), the $h_1$-mediated contributions to the EDMs are sub-leading. 
The $h_2$- and $h_3$-mediated Barr-Zee diagrams constitute the dominant contributions to EDMs. 
The cancellation between them leaves the blank region after imposing the recent and projected bounds of the eEDM, nEDM and diamagnetic aEDM experiments.
The CPV mixing angle is bounded to be $|s_{\alpha_b}|\lesssim\mO(10^{-3})$ with $t_\beta\sim 0.5$ .
Here, the eEDM cancellations happen around $t_\beta\sim 1$ and $t_\beta\sim 20$. A benchmark is given in Table.~\ref{tab:BP1} to illustrate the cancellation mechanism with $\Delta M=0.5\,\GeV$.
The right panel of Fig.~\ref{fig:BZ} (with $\Delta M=50$ GeV) depicts the EDM constraints with relatively larger mass splitting, which corresponds to the cancellation mechanism explored in Refs.~\cite{Shu:2013uua,Bian:2014zka,Inoue:2014nva}.
The corresponding CPV mixing angle is bounded to be $|s_{\alpha_b}|\lesssim\mO(10^{-4})$ by the projected $^{225}$Ra EDM experiment~\cite{Bishof:2016uqx}. 
In this situation, the constraint from the project $^{225}$Ra EDM measurement is the most stringent one since the cancellation between two heavy Higgs boson mediated Barr-Zee diagrams is negligible.
The comparison between the smaller and larger mass splittings indicates that the projected $^{225}$Ra EDM constraint can be relaxed with the decrease of $\Delta M$.

 \vspace*{0.5cm}
\begin{table}[!h]
\centering
\begin{tabular}{cccccccccccccccccccc}
\hline\hline
  $t_\beta$ &  $s_{\alpha_b}$ &  $s_{\alpha_c}$ &$\Delta M$(GeV)\\
\hline
   0.45&0.001&0.47 &0.5\\ [+1mm]
\hline
  $|d_e^{\rm tot}|$ &  $d^{h_1}_e $ & $d^{h_2}_e $   & $d^{h_3}_e$ \\
 \hline
 1.71$\times 10^{-30}$&9.37$\times 10^{-30}$&-3.10$\times 10^{-28}$&3.03$\times 10^{-28}$\\[+1mm]
\hline
$|d_{n}^{\rm tot}|$   &   $d^{h_1}_{n}$   &  $d^{h_2}_{n}$  &  $d^{h_3}_{n}$   \\
\hline
2.12$\times 10^{-28}$&-2.75$\times10^{-28}$&-3.43$\times10^{-26}$&3.44$\times10^{-26}$ \\[+1mm]
\hline 
$|d_{\rm Hg}^{\rm tot}|$  &  $d^{h_1}_{\rm Hg}$  & $d^{h_2}_{\rm Hg}$ & $d^{h_3}_{\rm Hg}$\\
\hline
3.76$\times 10^{-31}$&-1.63$\times 10^{-31}$&-3.00$\times 10^{-28}$&3.00$\times 10^{-28}$  \\[+1mm]
\hline
$|d_{\rm Ra}^{\rm tot}|$  &  $d^{h_1}_{\rm Ra}$  &  $d^{h_2}_{\rm Ra}$  &$d^{h_3}_{\rm Ra}$  \\
\hline
7.95$\times 10^{-29}$&3.97$\times10^{-28}$&-1.97$\times10^{-25}$&1.97$\times10^{-25}$\\[+1mm]
\hline
\hline
\end{tabular}
\def\baselinestretch{1.1}
\caption{
A benchmark in the CPV type-II 2HDM, with all other parameters set as in the left panel of Fig.~\ref{fig:BZ}. 
The individual contributions from each neutral Higgs boson $h_i$ and the total results are shown for eEDM, nEDM, $^{199}$Hg and $^{225}$Ra EDMs.
The unit for all EDMs is $e\cdot {\rm cm}$.}
\def\baselinestretch{1.0}
\label{tab:BP1}
\end{table}

We note that a relatively smaller $\Delta M$ gives rise to a relatively larger magnitude of CPV mixing angle $|s_{\alpha_c}|$, with the fixed $s_{\alpha_b}$~\footnote{The issue is indicated by Eq.~\ref{eq:Hmass_relation} and has been explored in Fig.~\ref{fig:pardec}.}. 
Therefore, the dominant contributions to EDM predictions are induced by $\alpha_c$ ($\alpha_b$) for a smaller (larger) mass splitting scenario.

\noindent{\bfseries MSSM}
In the MSSM, the CPV mixing term in the Higgs mass squared matrix is induced at loop level~\cite{Pilaftsis:1999qt}.
This further modifies the Higgs Yukawa couplings through the CPV vertices~\cite{Pilaftsis:1999qt,Carena:2001fw}.
For the nearly degenerate heavy Higgs bosons, the lightest SM-like Higgs almost decouples from the CPV mixing in the mass matrix of Eq.~\eqref{eq:Hmatrix}. 
Indeed, we find that $|\tilde c^h_f|\ll |c_f^h| $ by using ${\tt CPsuperH}$~\cite{Lee:2003nta, Lee:2012wa}~\footnote{This is consistent with the finding in Ref.~\cite{Li:2015yla}, the EDMs experiments almost preclude the CPV in the SM-like Higgs (the lightest one) in the CPV MSSM~\cite{Li:2015yla,Carena:2015uoe}.}.
Thus, the SM-like Higgs contributions to both e(q)EDM and CEDM through the Barr-Zee diagrams are negligible. 
Correspondingly, one can expect the dominant EDM contributions coming from the Barr-Zee diagrams mediated by the heavy Higgs bosons.

The CPV arising from the chargino sector embraces the high efficiency in generating the BAU via the electroweak baryogenesis (EWBG), as has been studied extensively in Refs.~\cite{CW:1997,Lee:2004we,Cirigliano:2009yd}~\footnote{We note that there were several relevant papers working on the line of this subject, pioneered by Carlos E.M. Wagner, J. Cline, M. Laine, K. Kainulaien, S. Huber, P. John, T. Prokopec, M. Schmidt, etc.}.
It also induces non-trivial contributions to the e(q)EDM via the $c_{f} \tilde c_{\gamma}$ and $\tilde c_{f} c_{\gamma}$ terms. 
The other important contributions to the e(q)EDM come from the stau sector through the $\tilde c_{f} c_{\gamma}$ term.
Therefore, one obtains~\footnote{We consider the heavy mass limit for the first and second generations of sfermions, which is necessary to accommodate the BAU~\cite{Cirigliano:2009yd}.
The one-loop EDMs are highly suppressed in the MSSM~\cite{Ellis:2008zy, Cirigliano:2009yd}.},
\beqn\label{MSSMedm}
d^E_{e,q} &\sim& \tilde{c}^A_{e,q}  \sum_{j=1,2}   \left( c^{\tilde{\chi}^\pm_{j}}_\gamma \ln( 1/z_{\tilde{\chi}^\pm_j}^A) + c^{\tilde{\tau}^\pm_{j}}_\gamma  \ln(1/z_{\tilde{\tau}^\pm_j}^A)  \right)
\nonumber\\
&&-  c^H_{e,q} \sum_{j=1,2} \tilde{c}^{\tilde{\chi}^\pm_{j}}_\gamma  \ln(1/z_{\tilde{\chi}^\pm_j}^H)\,,
\eeqn
with $z_x^y\equiv  M_x^2/M^2_y$, and the terms with superscripts of $A,H$ denoting the contributions of approximate $CP$-odd and $CP$-even Higgs bosons, and $c_\gamma^{\tilde{\chi}_i^\pm}\, , c_\gamma^{\tilde{\tau}_i}$ are given in Ref.~\cite{Bian:2014zka}.
Since $\tilde{c}^A_{\ell\,,d} \propto t_\beta $ and $c^H_{\ell\,,d} \propto 1/ c_\beta $, the CEDM also gets enhancements with large-$t_\beta$.The corresponding expression of $d^C_q$ could be obtained with $c_\gamma$ replaced by $c_g$, except for some form factors~\cite{Ellis:2008zy}.

\begin{figure}[htp]
 \includegraphics[width=6cm,height=4cm]{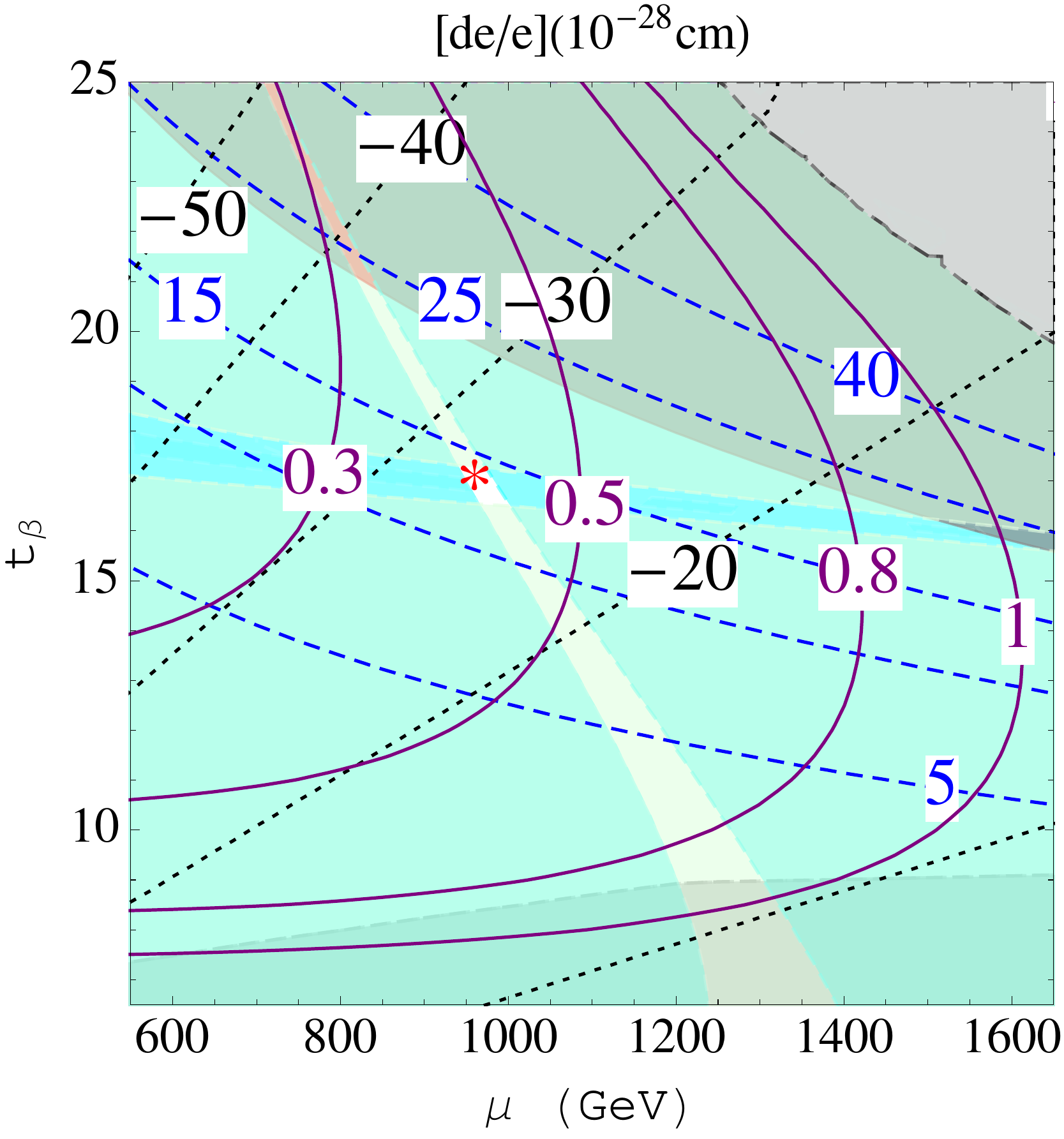}
\caption{ 
Parameter regions allowed by the EDM measurements on the $(\mu\,,t_\beta)$ plane, with color-code same as in Fig~\ref{fig:BZ}. 
The gray region (with a small $t_\beta$) is excluded since $M_1<124\,\GeV$. 
The chargino and stau contributions to the eEDM are plotted with black dashed and blue dashed contours, respectively.  
The mass splitting of $\Delta M$ is shown by purple contours . 
The parameters are set as in Ref.~\cite{Bian:2014zka} except that ${\rm Im}(\mu M_2^*)={\rm Im}(\mu A_f^*)=45^0$. }\label{fig:edmmutanbeta}
\end{figure}

By using the ${\tt CPsuperH}$~\cite{Lee:2003nta, Lee:2012wa} based on the consideration in Ref.~\cite{Bian:2014zka}~\footnote{In our scenario, the effects of the rainbow diagram contributions~\cite{Yamanaka:2012ia} are not important.}, we evaluate different EDMs numerically.  
As indicated in Fig.~\ref{fig:edmmutanbeta}, charginos\,(staus) lead to negative\,(positive) contributions to the eEDM, thus a cancellation in the eEDM.
The cancellation is highly related to the mass degeneracy of $\Delta M\sim\mO(0.1)-\mO(1)\,\GeV$, as plotted by the purple contours of Fig.~\ref{fig:edmmutanbeta}. 
The magnitude of $\Delta M$ increases with the increasing (decreasing) of $\mu$ ($t_\beta$). 
It depicts that a smaller $\Delta M$ leaves a narrower blank region allowed by EDM constraints. 
This is due to the fact that a smaller $\Delta M$ leads to a larger CPV phase, which is similar to the case of type-II 2HDM explored in the previous section.
The larger size of CPV in the heavy Higgs sector of MSSM with the degenerate mass is found for the first time.
The projected diamagnetic aEDM experiments might fill the blind points that were uncovered by the eEDM results from the ACME collaboration.
Again, the potential cancellation in the e(q)EDMs as indicated by Eq.~\eqref{MSSMedm} leaves the dominant contributions to nEDM and diamagnetic aEDMs from the CEDM and Weinberg operator couplings. 
Different from the 2HDM case, the cancellation effect in the CEDM is very small, due to the smallness of the loop-induced heavy Higgs mixings, as depicted by the purple contours of Fig.~\ref{fig:edmmutanbeta}. 
Furthermore, the contributions to the Weinberg operator from the heavy Higgs bosons cancel a lot in Eq.~\eqref{eq:CG}, as checked by ${\tt CPsuperH}$.
Instead, the gluino contributions are the dominant ones, and the details can be found, e.g., in Ref.~\cite{Ellis:2008zy}.
The projected $^{255}$Ra EDM measurements set $t_\beta\sim 16-18$ for 600 GeV$\leq\mu\leq$ 1600 GeV. 
A benchmark point (red star in Fig.~\ref{fig:edmmutanbeta}) is presented in Table.~\ref{tab:BP2}. 
We list the dominant contributions from charginos and staus to the eEDM, while other contributions from stops are sub-leading.

 \vspace*{0.5cm}
\begin{table}[!htp]
\centering
\begin{tabular}{cccccccccccccccccccc}
\hline \hline
  $t_\beta$ &$\mu$ (TeV) &$M_1$ (GeV)&$\Delta M$ (GeV) &
&\\
\hline
  17.5 &0.88 &125.6&0.4&&\\ [+1mm]
\hline
 $|d_e|$   & $d_e\, (\tilde{\tau})$   &   $d_e\,(\tilde{\chi}^\pm)$  && 
 &\\
 \hline
  6.8$\times 10^{-30}$&1.4$\times 10^{-27}$&-2.7$\times 10^{-27}$ &&\\[+1mm]
\hline
   $|d_n|$& $|d_{\rm Hg}|$ &$ | d_{\rm Ra}|$  & &\\
\hline
  4.8$\times 10^{-27}$&4.7$\times 10^{-30}$  &   0.5$\times 10^{-28}$& &\\[+1mm]
\hline \hline
\end{tabular}
\def\baselinestretch{1.1}
\caption{A benchmark in the MSSM, with EDMs in units of $e\cdot {\rm cm}$ and the other parameters set as Fig.\ref{fig:edmmutanbeta}.}
\def\baselinestretch{1.0}
\label{tab:BP2}
\end{table}


\noindent{\bfseries Conclusions and discussions}
\label{section:conclusion}
In this paper, we observe a general cancellation mechanism in the calculations of eEDM, nEDM, and diamagnetic aEDMs due to the cancellation through the $h_i$-mediated eEDM, q(C)EDM operators and the Weinberg operators. 
The cancellation of eEDM could happen in larger mass splitting of heavy Higgs scenario,  
in this situation the cancellation of eEDM could happen between different degree freedom (as observed in the framework of CPV type-II 2HDM by Ref.~\cite{Bian:2014zka}), where a larger CPV could exist in the SM-like Higgs. 
A smaller mass splitting leads to a larger size of CPV beyond the SM like Higgs sectors, which opens the opportunity to explain BAU via the EWBG. 
The chance to evade a relatively larger CPV phase by further probing the nEDM, the current ACME and recent diamagnetic aEDMs, and even the future $^{225}$Ra EDM experiment is found, which is characterized by the mass degeneracies of the heavy Higgs bosons.
The typical magnitude is found to be $\Delta M\sim \mO(10^{-1})- \mO(1)$ GeV. 
The CPV type-II 2HDM and MSSM have been investigated to reveal such features.
In general, one expects the cancellation mechanism happens in NP models with CPV mixings in the Higgs sector.

The projected $^{225}$Ra EDM measurements and the high-luminosity LHC experiments are likely to detect or fully exclude the scenarios.
The polarization of top quark might also help to identify the CP properties of the two heavy Higgs bosons~\cite{Carena:2016npr}, depending on the CPV magnitude and mass splitting between two heavy Higgs bosons. 
The observation of the CPV at $\mu^+\mu^-$ collider~\cite{Pilaftsis:1996ac} can be used to search for sizable CPV mixing in the SM-like Higgs boson, which corresponds to the case with larger mass splitting between heavy Higgs bosons.
The analysis of the spins of the $\tau$ decay final state at the $\gamma \gamma$ colliders and the LHC~\cite{Arbey:2014msa} can be promising for the CPV heavy degenerate Higgs scenarios. 
Furthermore, the triple Higgs self couplings can be constrained by the CPV mixings in such models~\cite{Bian:2016awe}.
We expect that the Higgs pair productions at the future electron-positron and hadron collider may detect the CPV effects indirectly, and discriminate two cancellation mechanisms.

\noindent{\bfseries Acknowledgements}
We would like to thank Carlos E.M. Wagner for reading our manuscript and providing valuable comments and suggestions.
This work is supported by National Science Foundation of China (under Grant No.\ 11605016 and No.\ 11575176), and the Fundamental Research Funds for the Central Universities (under Grant No.\ WK2030040069). 
We would like to thank the Kavli Institute for Theoretical Physics China at the Chinese Academy of Sciences for their hospitalities when part of this work was prepared.


\appendix
\section{Anatomy of fEDM ($f=e,q$), CEDM, and Weinberg operator in the nEDM and diamagnetic aEDMs}
\label{sec:CPsuperH}

In order to make a more transparent picture of the cancellation mechanism in deriving the nEDM and diamagnetic aEDMs, we take the CPV type-II 2HDM as an example. 
We first investigate the relation between the mass splitting of heavy Higgs bosons and CPV mixings. 
Afterwards, we explore the individual contributions to the nEDM and diamagnetic aEDMs. 
\\

{\bf (1) CPV versus mass splitting of heavy Higgs bosons } \\

\begin{figure}[htb]
\includegraphics[width=4.cm,height=3.cm]{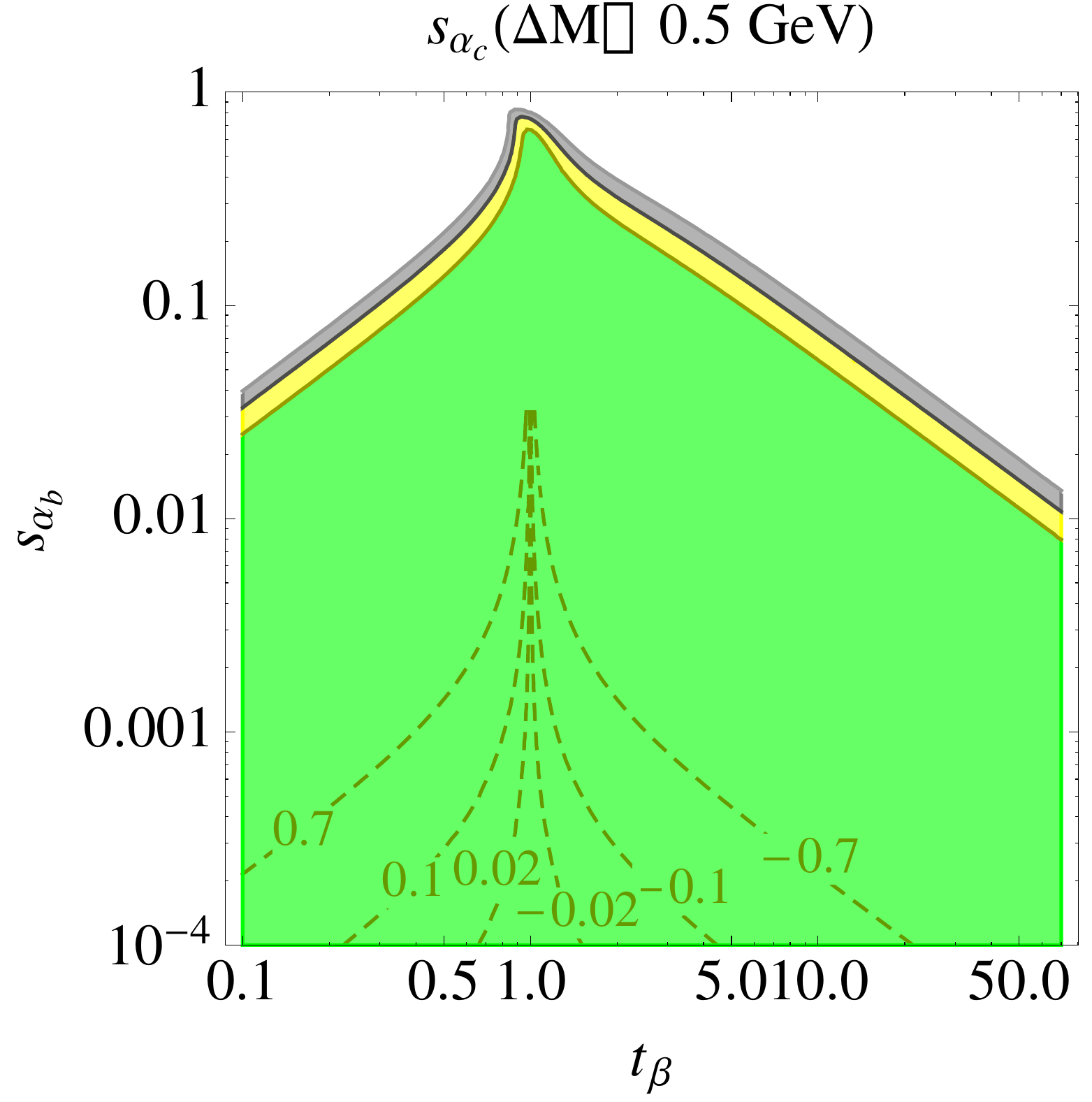}
\includegraphics[width=4.cm,height=3.cm]{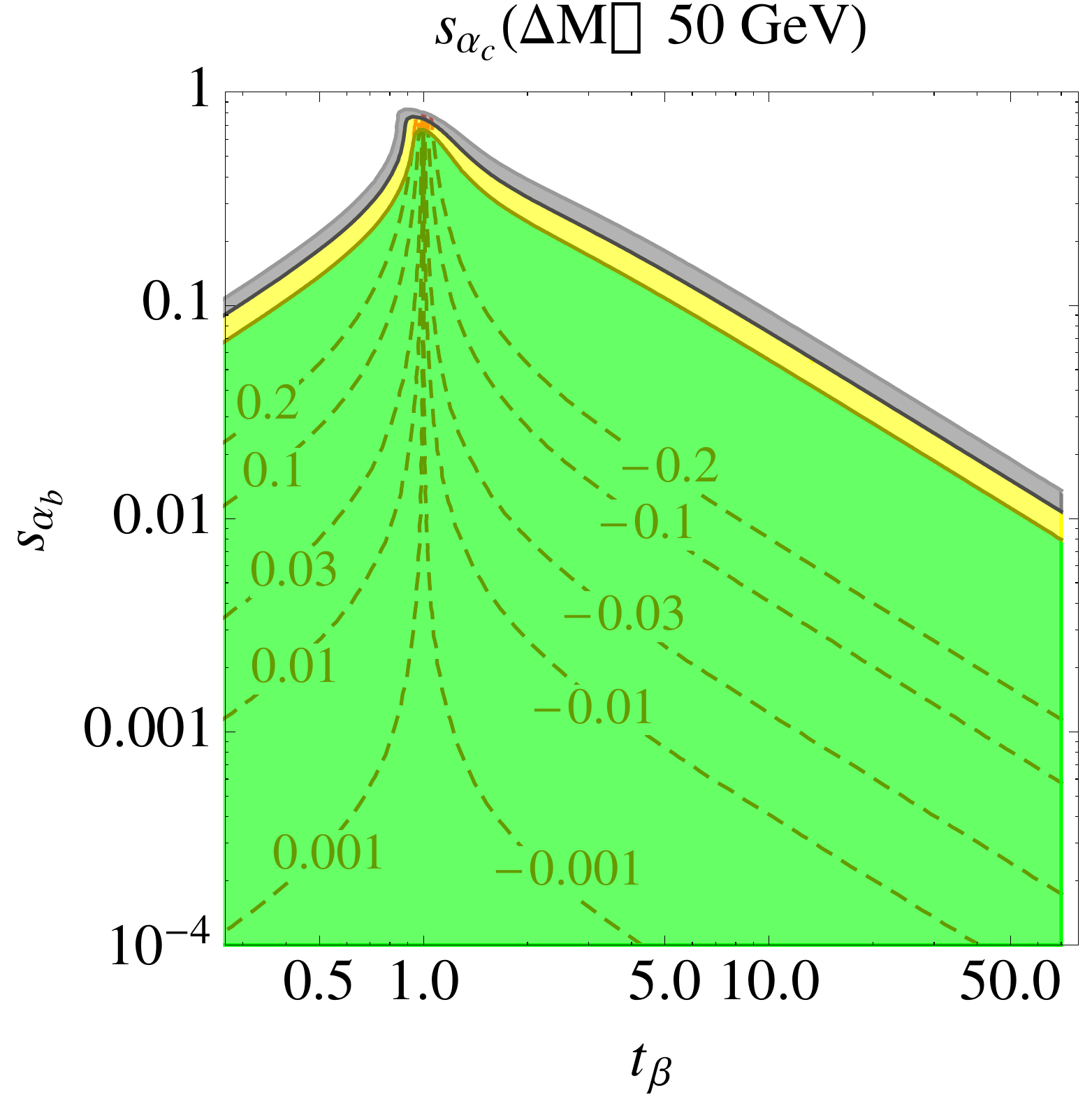}
\includegraphics[width=3.9cm,height=3.cm]{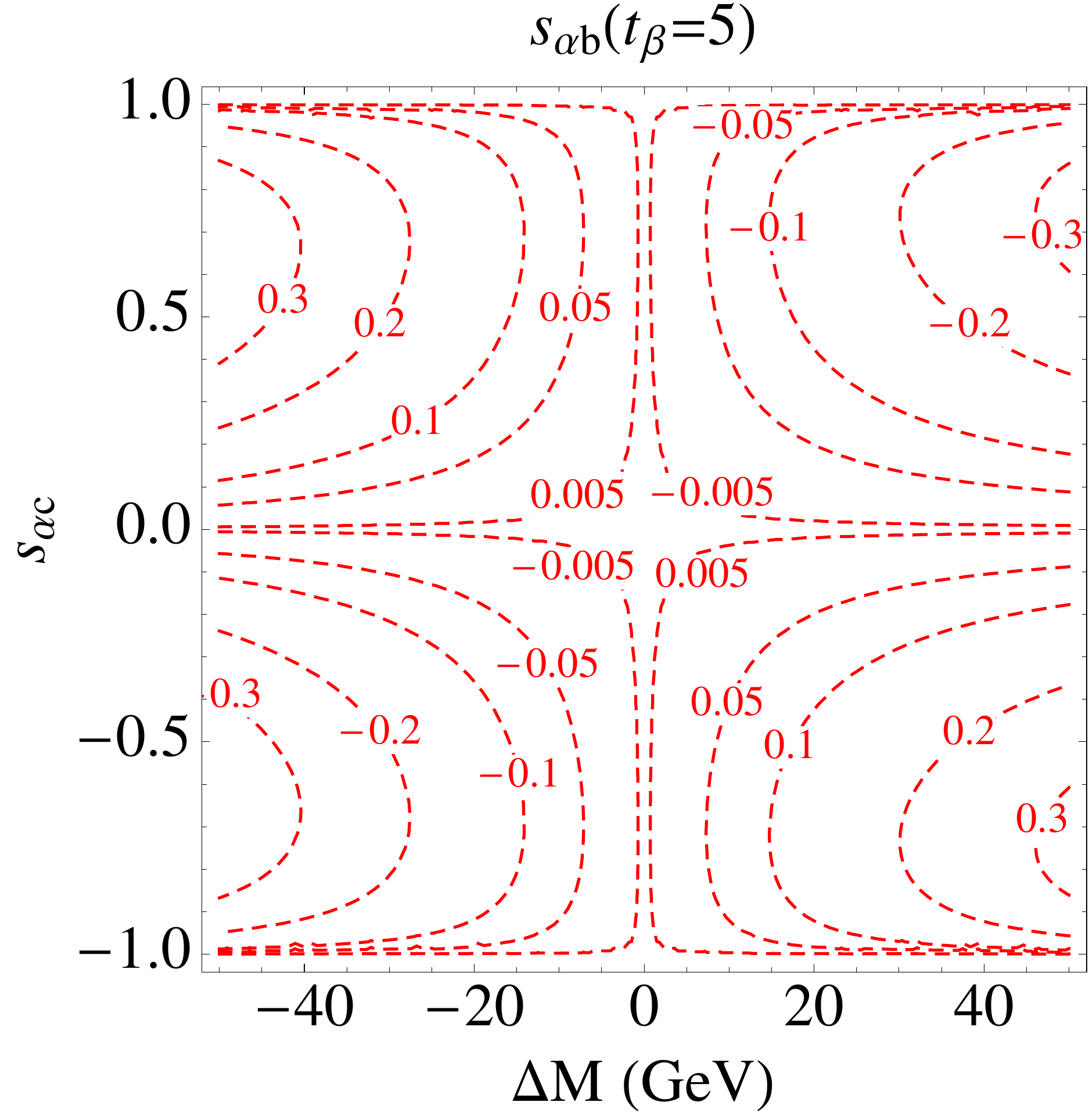}
\includegraphics[width=3.9cm,height=3.cm]{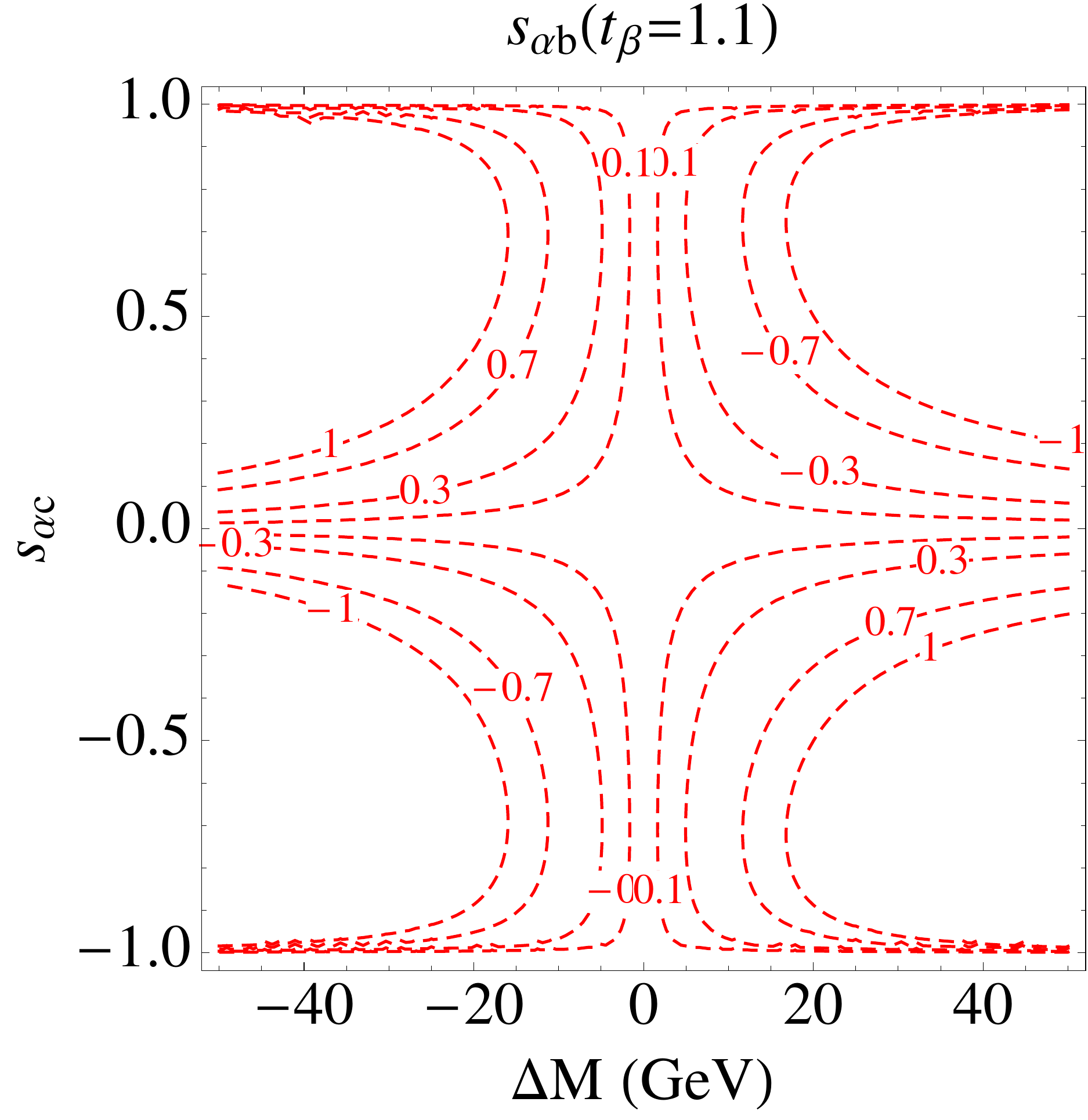}
\caption{\label{fig:pardec}
The contours of $s_{\alpha_b}$ and $s_{\alpha_c}$ as a function of $(\Delta M, s_{\alpha_c})$ and $(t_\beta\,, s_{\alpha_b})$(in the alignment limit) for top and bottom panels.  
The contours of $s_{\alpha_c}$ (dashed lines) as a function of $(t_\beta\,, s_{\alpha_b})$ in the alignment limit of $\beta-\alpha=\pi/2$.
The grey, yellow, and green regions are the $1\,,2\,,3\,\sigma$ allowed regions of the LHC $7\oplus 8\,\TeV$ Higgs data~\cite{hfit}. }
\end{figure}

The relation between two CPV mixing angles is given by Eq.~(6).
To visualize the relation between the CPV magnitude and mass splitting of heavy Higgs bosons, the Fig.~\ref{fig:pardec} is shown by using the parameters as in {\it Fig.2 of the main body of the paper}.

With larger deviation of $t_\beta$ from 1 and smaller mass splitting of $\Delta M $, one gets larger $|\alpha_c|$ in comparison with $\alpha_b$, as depicted by two top panels of Fig.~\ref{fig:pardec}. 
In the bottom panels, we fix $t_\beta$ and plot $s_{\alpha_b}$ as a function of $s_{\alpha_c}$ and $\Delta M$.
The comparison of two bottom panels demonstrates that a bigger $t_\beta$ leads to a smaller $s_{\alpha_c}$ for a fixed $\Delta M$, and the magnitude of $s_{\alpha_b}$ increases with the larger inputs of $\Delta M$ for a given $t_\beta$. 
\\

{\bf (2) CPV versus nEDM and aEDMs } \\

The nEDM and diamagnetic aEDMs are related to the Wilson coefficients~\cite{Bian:2014zka,Engel:2013lsa,Inoue:2014nva} through the definitions of
$\delta_f \equiv -\Lambda^2 d_f^E/(2 e Q_f m_f)\,, \tilde \delta_q \equiv -\Lambda^2 d_q^C/(2 m_q)\,, C_{\tilde G} \equiv \Lambda^2 d^G/3 g_s\,,
$ with $m_f$ and  $\Lambda$ being the fermion masses and the NP scale, respectively~\footnote{Here, it should be noted that theoretical calculations of the nEDM and the diamagnetic aEDMs suffer from uncertainty of nuclear and hadronic matrix elements~\cite{Jung:2013mg,Engel:2013lsa}.}. 

\begin{figure}[htb]
\centering
\includegraphics[width=4.1cm,height=3.8cm]{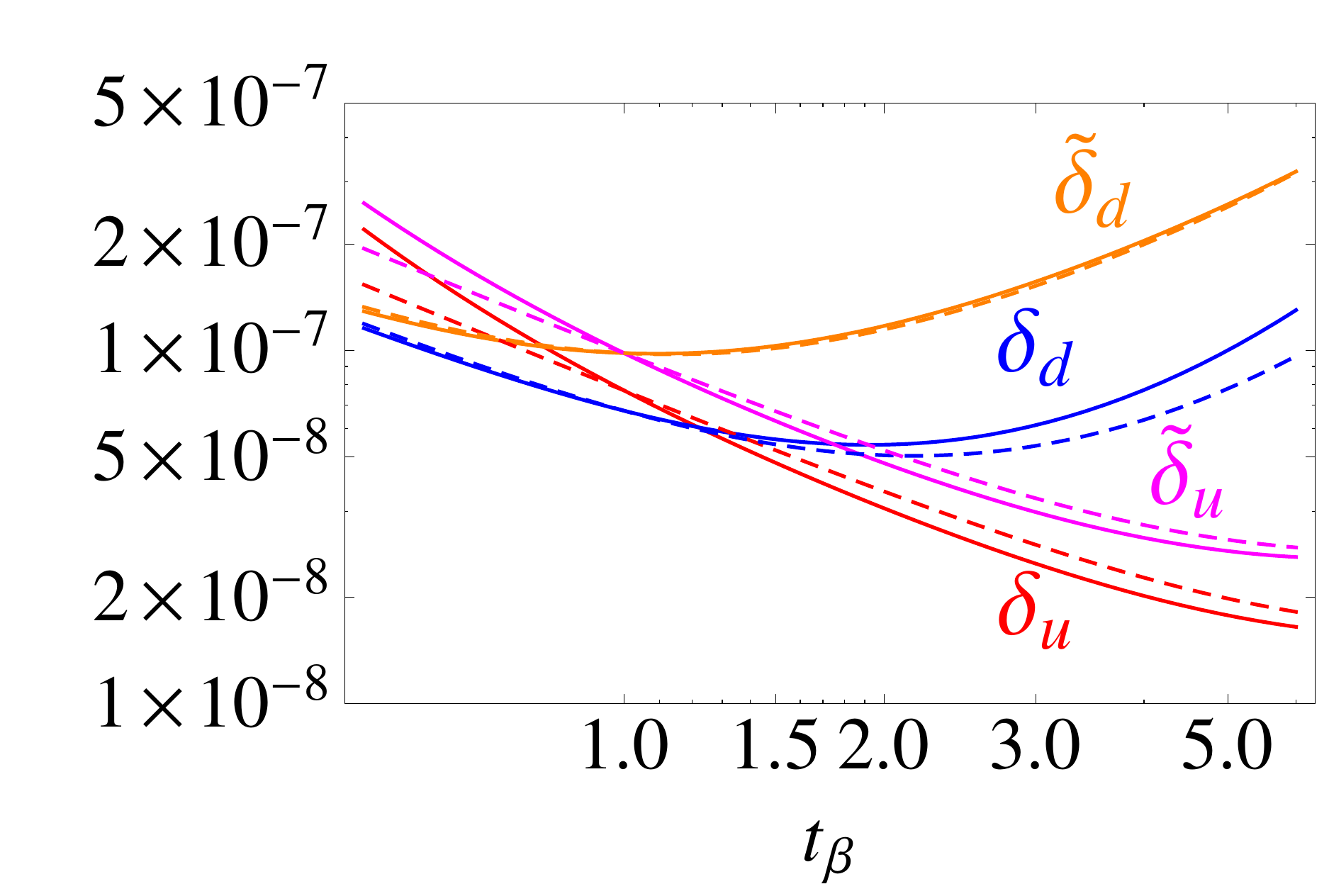}
\includegraphics[width=4.1cm,height=3.8cm]{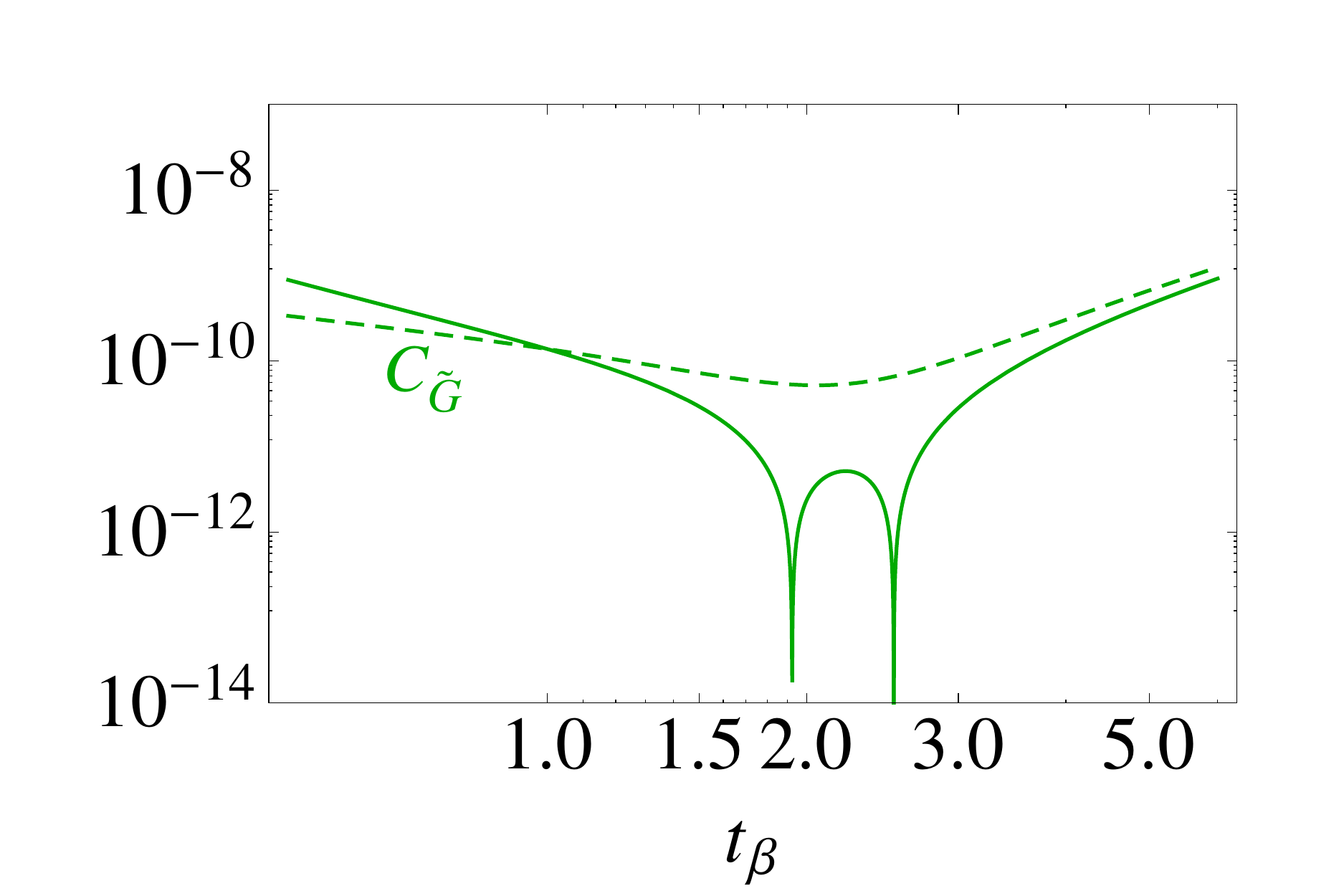}
\caption{\label{fig:difcon}
Wilson coefficients of q(C)EDM (left) and Weinberg operator (right) versus $t_\beta$ at the hadron scale of $\mO(1)\,\GeV$. 
The parameter inputs are $M_3\,(m_{\rm soft})=500\,(300)\,\GeV$, with solid (dashed) lines represent the case of $\Delta M$ being $0.5\,$(50)$ \GeV$.
 }
\end{figure}

To make the cancellation of nEDM and diamagnetic aEDMs more transparent, we take $s_{\alpha_b}=0.001$ as a benchmark. 
The variations of Wilson coefficients for q(C)EDM and Weinberg operator with different mass splittings of two heavy Higgs bosons are displayed in Fig.~\ref{fig:difcon}.
It is clear that a smaller mass splitting leads to stronger cancellations to the Wilson coefficients for q(C)EDM and Weinberg operators aroud $t_\beta\sim1$.
This further leads to the blank regions in two panels of {\it Fig.~2 }~\footnote{It should be mentioned that the effects of $m_{\rm soft}$ is tiny, since it only affects the subdominant Barr-Zee diagrams in e(q)EDM. 
For explicit formula, we refer to Ref.~\cite{Inoue:2014nva}. 
The effects of the heavy Higgs boson masses are negligible because they enter into e(q)EDM, CEDM and $C_G$ through Eq.~3 and Eq.~4 of the main body the paper, and are logrithmically suppressed.}.

\section{The implementation of EWBG in 2HDM and MSSM}

In the type-II 2HDM, the heavy neutral Higgs bosons (with masses $\geq 400$ GeV) make the SFOPT feasible~\cite{Fromme:2006cm,Dorsch:2013wja}.
During the SFOPT, the CPV sources from top quark~\cite{Fromme:2006cm,Shu:2013uua} (or bottom quark~\cite{Liu:2011jh}) inside the nucleate bubble wall generate the CPV charge asymmetries, and are converted into BAU by the electroweak sphaleron transition process. 
The relative phase $\xi$ between two Higgs doublets is related to the CPV source, while the CPV mixing angle $\alpha_c$ is highly related to the EDM predictions.
Their relations can be found in Eqs.~(2.11) of Ref.~\cite{Bian:2016awe}.

The magnitude of the CPV phase explored in this work is about half of the one in Ref.~\cite{Bian:2014zka}.
To make the gaugino-higgsino CPV sources drive the EWBG~\cite{Lee:2004we,Cirigliano:2009yd}, we need a higher wall velocity $v_w$ or thinner wall (lower magnitude of wall width $L_w$).
This can be derived during the SFOPT, depending on detail of the thermal theory of the universe. 
The SFOPT may still be available with a light stop being allowed by the recent studies based on LHC run I~\cite{Liebler:2015ddv}, for which we left to the future study.

It should be mentioned that, the CPV in the next to minimal supersymmetry Standard model (NMSSM) might also embrace such property. 
One can expect its behavior similar to the 2HDM case, since the CPV mainly arises from the tree level rather than from the radiative corrections.
The details will be studied in Ref.~\cite{Bian}. 
Meanwhile, it's feasible to realize SFOPT and implement EWBG mechanism in the CPV NMSSM, as studied in Ref.~\cite{Bi:2015qva,Huang:2014ifa,Kozaczuk:2014kva}.

\end{document}